\documentclass[12pt,letterpape]{article}
\usepackage[a4paper, total={7in, 10in}]{geometry}

\usepackage{graphicx}
\usepackage{helvet}
\usepackage{authblk}
\usepackage{hyperref}
\usepackage{amsmath} 
\usepackage{amssymb} 
\usepackage{orcidlink} 
\usepackage[super,comma,sort&compress]  
   {natbib}

\usepackage[T1]{fontenc}    % use 8-bit T1 fonts
\usepackage{hyperref}       % hyperlinks
\usepackage{url}            % simple URL typesetting
\usepackage{booktabs}       % professional-quality tables
\usepackage{amsfonts}       % blackboard math symbols
\usepackage{nicefrac}       % compact symbols for 1/2, etc.
\usepackage{microtype}      % microtypography
\usepackage{lipsum}
\usepackage[table]{xcolor}
\usepackage{comment}
\usepackage{float}

\newcommand{\be}{\begin{equation}}
\newcommand{\ee}{\end{equation}}
\newcommand{\ba}{\begin{align}}
\newcommand{\ea}{\end{align}}
\newcommand{\bea}{\begin{eqnarray}}
\newcommand{\eea}{\end{eqnarray}}

\usepackage{amsmath}

\newfloat{Box}{tbp}{lob}
\definecolor{boxbg}{rgb}{0.9,0.9,0.9}
\newcounter{theoryboxc}
\newsavebox{\savetheorybox}
\newcommand{\printtheorybox}[1]{\fcolorbox{black}{boxbg}{#1}}
\newenvironment{theorybox}[1][]%
{\begin{Box}\begin{lrbox}{\savetheorybox}%
 \begin{minipage}{0.985\linewidth}% \compacteqs%
 \refstepcounter{theoryboxc}%
 \textbf{Box \thetheoryboxc. #1} %
}
{\end{minipage}%
 \end{lrbox}%
 \printtheorybox{\usebox{\savetheorybox}}%
 \end{Box}
}

\makeatletter
\renewcommand{\maketitle}{\bgroup\setlength{\parindent}{0pt}
\begin{flushleft}
  \textbf{\@title}
  
  \@author
\end{flushleft}\egroup}
\makeatother

\title{How Intrinsic Motivation Underlies Embodied Open-Ended Behavior}
\date{}
\author[1-3,*]{Rubén Moreno-Bote}
\author[4]{Ralf Haefner}
\author[5]{Jordi Galiano-Landeira}
\author[6]{Tianming Yang}
\author[7,8]{Pedro Maldonado}

\affil[1]{Center for Brain and Cognition, Universitat Pompeu Fabra, 08002, Barcelona, Spain}
\affil[2]{Department of Engineering, Universitat Pompeu Fabra, 08002, Barcelona, Spain}
\affil[3]{Serra Húnter Fellow Programme, Universitat Pompeu Fabra, Barcelona, 08002, Spain}
\affil[4]{Department of Brain and Cognitive Sciences and Center for Visual Science, University of Rochester, Rochester, New York, New York, United States of America}
\affil[5]{Centro Internacional de Neurociencia y Ética (CINET), Madrid 28010, Spain}
\affil[6]{Institute of Neuroscience, Key Laboratory of Brain Cognition and Brain-inspired Intelligence Technology, Center for Excellence in Brain Science and Intelligence Technology, Chinese Academy of Sciences, Shanghai 200031, China}
\affil[7]{Deparment of Neuroscience, Universidad de Chile, Santiago, Chile}
\affil[8]{National Center for Artificial Intelligence, CENIA, Santiago, Chile}

\affil[*]{Correspondence: ruben.moreno@upf.edu}

\begin{document}

%Target journal: Trends in Cognitive Sciences (word limit 4000, not including boxes, tables, captions, abstract), Neuroscience \& Biobehavioral Reviews, others?

\maketitle

\subsection*{Abstract}
Although most theories posit that natural behavior can be explained as maximizing some form of extrinsic reward, often called utility, some behaviors appear to be reward independent.
For instance, spontaneous motor babbling in human newborns and curiosity in little kids and other animals seem to elude a simple explanation in terms of extrinsic reward maximization.
Rooted in these observations, intrinsic motivation has emerged as a potentially major driver of behavior.
However, only recently have several quantitative and foundational theories of intrinsic motivation been put forward.
We first provide a general framework to understand behavior as being organized hierarchically: objective--intrinsic reward, or motivation--drives, goals and extrinsic reward.
We next review the main formalizations of intrinsic motivation, including empowerment, the free energy principle, information-gain maximization, and the maximum occupancy principle. 
These theories produce complex behavior by promoting, in various ways, entropic action-state paths. 
The presence of a single intrinsic motivation objective breaks infinite regress, as drives and goals act only temporarily to serve the objective. 
Extrinsic rewards, such as sugar or protein, are just a means to achieve the objective. 
Bounded cognition and embodiment impose constraints and boundary conditions for the intrinsic motivation objective. 
By virtue of their capability to generate complex behavior in a task-agnostic manner,
theories of intrinsic motivation promise to become successful generative models of open-ended, embodied behavior.

\subsection*{Highlights}

\begin{itemize}

\item Neuroscience and behavioral sciences posit that behavior is primarily driven by extrinsic reward--or utility--maximization

\item There are reward-independent and task-agnostic behaviors that seem to be intrinsically motivated

\item Intrinsic motivation can be formalized as the maximization of action-state path entropy, or variations thereof

\item Extrinsic rewards are a means for achieving the maximization of the intrinsic motivation objective

\item Embodiment and bounded cognition constrain and shape the behavior generated by intrinsic motivation 

\end{itemize}

\subsection*{Glossary}

\begin{itemize}

\item \textbf{Behavioral Policy}: In a Markov Decision Process (MDP), a behavioral policy is the probability of taking action $a$ when in state $s$, denoted as $\pi(a|s)$. An MDP is additionally defined by an extrinsic reward function $r(s,a)$, a set of actions available in each state, and a state transition probability $p(s'|s,a)$, which describes the probability of transitioning to state $s'$ when performing action $a$ in state $s$. The policy can change in the course of learning, while the extrinsic reward function, to be maximized by the agent, is assumed to remain fixed.  

\item \textbf{Bounded Cognition}: Neural processes serving a goal, all of which are naturally bounded by the limits of the brain machinery. Examples are our tight limitations in working memory and planning breadth and depth \cite{mastrogiuseppe2022deep,moreno2020heuristics,van2023expertise,callaway2024revealing}. 

\item \textbf{Embodiment}: Integration of the nervous system with a complex body that should be operating in a closed-loop manner across multiple spatiotemporal scales (cells, milliseconds; bodily behaviors, seconds to years).
An isolated brain in a dish, lacking embodiment, arguably holds little relevance to understanding cognition and intelligence in general.  

\item \textbf{Homungoalus}: By analogy to the homunculus problem, the problem of what goal explains our current goal, and what goal is on top of the former, and so on, falling in an infinite regress \cite{shenhav2024affective}.  

\item \textbf{Objective, Drives and Goals}: (see Fig. \ref{fig:fig1}) An objective is an overarching principle of behavior, which lies at the top of the hierarchy (e.g., exploration, empowerment, survival). The objective is defined as unique and immutable, and can be implemented as maximization of an intrinsic reward signal (intrinsic motivation). 
A drive is a temporary biological need or impulse (e.g., thirst) at the next level of the hierarchy that is used to aid the objective (e.g., thirst to keep exploration or promote survival). Drives can compete. A goal, lowest in the hierarchy, is a plan with an ending state to satisfy a (temporary) drive (e.g., deciding to seek water and choosing one of two nearby wells to satisfy thirst). In the absence of drives, goals still emerge to aid in the objective (e.g., movement with guided randomness to aid empowerment or occupancy).

\end{itemize}

\newpage

\subsection*{Extrinsic Reward Maximization vs. Intrinsic Motivation}

While most theories of behavior assume that animals are extrinsic reward maximizers, theoretical and experimental considerations challenge this hypothesis.
Think of a robot the way an engineer would: you want to design the robot to perform a specific task, such as washing dishes and putting them in the rack.
As an engineer, you are free to choose the extrinsic reward function (Box 1). Which one would you use? The robot could be designed to maximize the number of cleaned and ordered dishes per unit of time, but only if they are not broken. But you can now define a new reward function where cleaning delicate glassware or heavy pots is more rewarding. On top of that, you can ask the robot to be energetically efficient and noiseless. At this point, you are left with an arbitrary choice of weights for each of the features, altogether defining the extrinsic reward function \cite{Eschmann2021}.
And what happens if there are suddenly no delicate glassware to clean up any more because the robot has been relocated to a university students' canteen? 
As an engineer, you would maybe like to change the reward function at this point, but it might be too late. 
The result for each of your extrinsic reward designs would be an inflexible behavior that becomes more deterministic as learning progresses: the optimal behavioral policy* (see Glossary) becomes a one-to-one mapping between states and actions in a Markov Decision Process \cite{sutton1998reinforcement}, that is, a deterministic policy. 
%(Fig. \ref{fig:fig2}b). 
%The behavior would be tailored to the shape of the extrinsic reward, and nothing else.

While appealing in some controlled situations, this scenario of extrinsic reward maximization falls short of generating realistic behavior. Indeed, when observing animals, once behavior such as feeding, drinking, or seeking shelter has been satisfied, they exhibit curiosity and exploration. For instance, human newborns motor-babble spontaneously without any clear objective 
\cite{adolph2015physical,petitto1991babbling}, kids explore with curiosity in both known and novel environments \cite{kidd2015psychology,gottlieb2013information} and prefer playing to harder games when playing for fun \cite{rule2023fun}, and adults continue to explore despite diminishing returns and increasing costs \cite{modirshanechi2025novelty}.
Collapse into repetitive behavior is rarely observed, and rather it is typically considered a signature of mental or psychological disorders, such as eating disorders \cite{agh2016systematic,rash2016review}.
Recent experimental results in various non-human animal models show that 
monkeys demonstrate a willingness to forgo rewards in order to obtain information that had no associated rewards \cite{blanchard2015monkeys},
rats prefer exploring mazes over staying in a home cage with freely available food \cite{rosenberg2021mice}, and zebrafish continuously move despite the fact that every location is similarly rewarding and there are nearby preys \cite{johnson2020probabilistic}.
If anything, it appears that extrinsic rewards are sought temporarily, but after enough consumption of them, what drives behavior is a signal of a different nature, e.g., an insatiable intrinsic motivation for exploration \cite{white1959motivation}. 
In this review, we address the question of what neuronal signals can drive behavior beyond extrinsic rewards, how to formalize theories of behavior that are based on intrinsic motivation signals using recent theoretical developments \cite{jung2011empowerment,ramirez2024complex,dacosta2023reward}, and how to reinterpret natural behavior by providing it with open-endedness, constrained only by bounded cognition* and embodiment*.

\subsection*{Hierarchies of drives and goals, but under one objective}

\begin{figure}[htbp]
\centering
\includegraphics[width=0.9\textwidth]{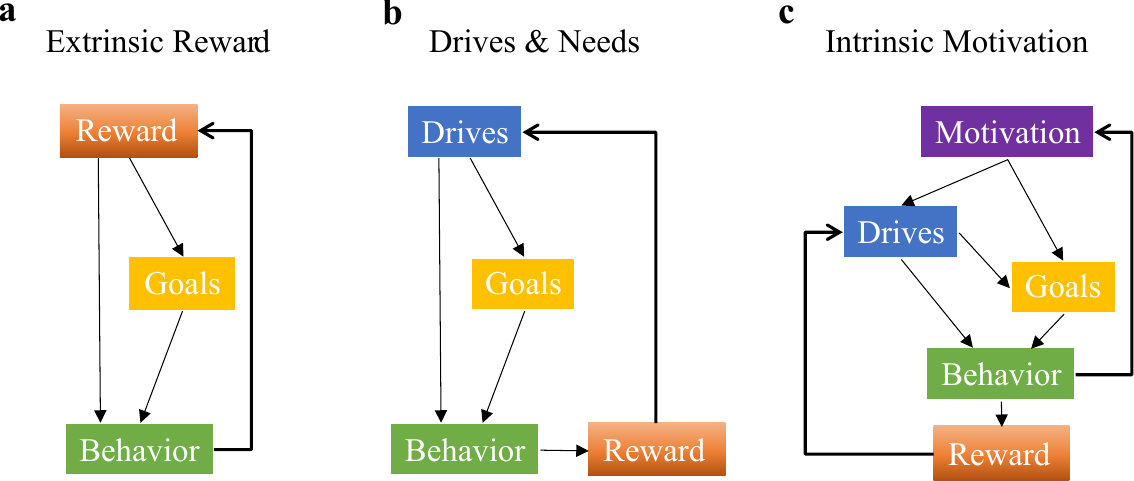}
\caption{
Hierarchical organization of objectives, drives, goals and extrinsic rewards in three global theories of behavior.
(a) Extrinsic reward maximization \cite{vonneumann2007theory,kahneman2013prospect,sutton1998reinforcement} posits an extrinsic reward function (denoted as reward in the figure) as the objective of the behavior. Extrinsic reward signals can directly trigger behavior, such as eating, but often goal-directedness, involving planning, is needed to obtain distant or future extrinsic rewards.
(b) Drive theory \cite{berridge2004motivation} of behavior places drives and needs at the top of the hierarchy \cite{rosado2022drive}. Drives compete between themselves to directly generate behavior or form temporary goals to satisfy them. Drives and needs are only satisfied if appropriate extrinsic rewards are gathered (e.g., water for thirst). 
(c) Intrinsic motivation theories \cite{jung2011empowerment,ramirez2024complex,pathak2017curiosity} put a single objective at the top of the hierarchy. The objective can be occupying action-state path space (Eq. \ref{eq:objective}). Generally, the objective consists in maximizing future cumulative intrinsic reward. This objective can generate drives and goals depending on the proximity of survival needs, which together generate behavior. Extrinsic rewards are obtained as a byproduct of behavior serving to the satisfaction of drives and needs, but they do not directly affect the objective.
The intrinsic reward mostly depends on behavioral state transition probabilities and the agent's policy.
Intrinsic and extrinsic rewards, needs and goals, and states and actions, are all {\em internal} to the agent, but they can represent the external reality, albeit in a potentially distorted and biased manner.  
}
\label{fig:fig1}
\end{figure}

\begin{figure}[htbp]
\centering
\includegraphics[width=0.75\textwidth]{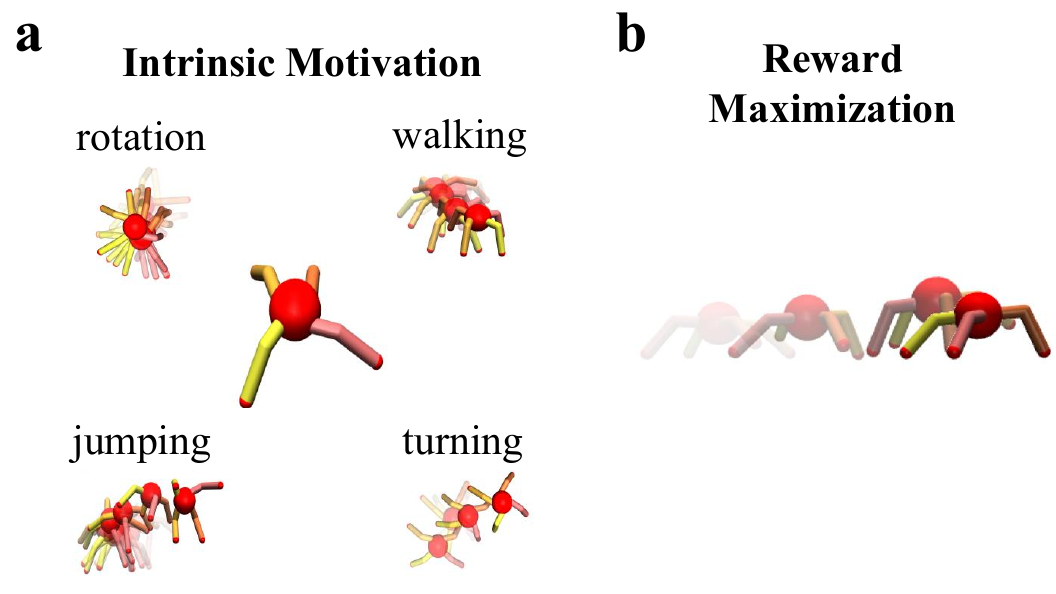}
\caption{
Intrinsic motivation versus extrinsic reward maximization in locomotion pattern generation in an embodied agent. The intrinsically motivated ant (panel a) discovers diverse behaviors, including walking and jumping. 
This diversity arises because the agent is intrinsically motivated to occupy action-state path space (Eq. \ref{eq:objective}), pushing it to visit the full range of behaviors possible within its physical constraints.
%This is because it tries to occupy action-state path space (Eq. \ref{eq:objective}), and therefore it is intrinsically motivated to explore all behaviors it is capable of, within its own physical constraints.  
In contrast, the extrinsic reward maximizer (panel b) is driven solely by positive rewards proportional to rightward speed. Over time, its behavior simplifies, converging to a single strategy: running to the right (\href{https://drive.google.com/file/d/1fqYDU8GRyppuSssb3vOhFVrlig_7hhAa/view?usp=drive_link}{Video 2}). For both agents, a terminal state is reached when the ant’s torso touches the ground (indicating a fall).
This terminal state introduces a natural boundary that limits and shapes behavior.
The virtual agent used here is the Ant in the MuJoCo environment \cite{todorov2012mujoco,ramirez2024complex}.}
\label{fig:fig2}
\end{figure}

While the intrinsically motivated behaviors described above seem to elude a simple explanation in terms of extrinsic reward seeking,
most current theories of behavior assume that extrinsic reward maximization is the objective* at the top of the hierarchy, fully defining how behavior is generated \cite{bogacz2006physics,drugowitsch2012cost,chen2025rational,rosado2022drive} (Fig. \ref{fig:fig1}a). 
Formally, an extrinsic reward function is a function of the agent's state and action(Box 1), and might reflect pleasure or satisfaction upon achieving states for which the function has high values.
Extrinsic reward maximization is a non myopic objective: the agent can forgo immediate reward, e.g., by engaging in learning or exploration, if this behavior predicts even higher extrinsic reward in the future. 
Indeed, extrinsic reward maximization is typically formalized as the maximization of discounted future cumulative extrinsic reward \cite{sutton1998reinforcement} (Box 1).
Below this top level there are temporary goals aimed at reaching intermediate milestones so that extrinsic reward is maximized in the long run.
For instance, intrinsic motivation to explore or to express curiosity are just a means to accomplish the extrinsic reward maximization objective; these intrinsic motivation can be subsumed into the notion of short-term goals.
Although attractive, extrinsic reward maximization predicts that behavior collapses to a small repertoire of patterns \cite{sutton1998reinforcement} (Fig. \ref{fig:fig2}b). 
Secondly, this theory does not specify what the extrinsic reward function is, leaving it as a matter of experimental investigation how to discover it through, e.g., inverse reinforcement learning \cite{drugowitsch2012cost,schultheis2021inverse,straub2022putting}, a problem that is generally intractable. 

To avoid behavioral collapse in extrinsic reward maximization, one can conceive a set of extrinsic reward functions that are either hierarchically organized or compete at the same level. Each of these reward functions might only be active temporarily, such as maximizing food intake or money. Therefore, multiple behaviors would be generated. 
However, this complicates the problem because then more unknown extrinsic reward functions have to be determined.  
In addition, the exercise of establishing a hierarchy of reward functions might lead to the homungoalus* problem \cite{shenhav2024affective}: which extrinsic reward function determines the activation of lower-level functions, which higher-level function determines which function is active at the top, and so on, falling in an infinite regress.
%at the top decides what other reward functions are active at the bottom, that sets  at a higher level makes active the current reward function at the lower level, and what other reward function makes active the one at the higher level, and so on.  

To account for behavior diversity and to avoid infinite regress, there must be an overarching objective from where all drives* and goals* emerge in a dynamic manner. We term this overarching objective of behavior intrinsic motivation, or intrinsic reward.
In this picture, intrinsic motivation is at the top of the hierarchy (Fig. \ref{fig:fig1}c). 
For instance, a central tenet in recent theories of intrinsic motivation is the idea of generating a diversity of states and actions \cite{ramirez2024complex,jung2011empowerment,park2023metra,sharma2019dynamics,gruaz2025merits} -- the persistence of energy spreading and movement generation are the simplest and arguably the most fundamental properties of living systems. Another overarching objective could simply be survival --which might also be directly related to the movement generation just described. 
%The differences is that movement can be grades, while survival is binary. 
%The intrinsic motivation is permanently active, in the sense that behavior should be guided to maximize it. 
%Intrinsic motivation is an open-ended objective because it is continuously active: it ought to be maximized on the long run and persistently. 

In this view, goals and drives lie right below the intrinsic motivation objective in the hierarchy: they produce temporarily active behaviors aiming at maximizing the intrinsic motivation objective.
According to this, primary (extrinsic) rewards are not the objective of behavior, but a means to maximizing intrinsic motivation -- acquiring enough water ensures that the animal can keep optimizing its ultimate objective, different from maximizing water intake itself. 
Intrinsic motivation theories solve the homungoalus problem because goals and drives are temporary active and exist only to maximize intrinsic motivation: a goal can be accomplished and a drive can be extinguished, but intrinsic motivation maximization is a persistent objective under which all behavior centers around.
For instance, maximizing the intrinsic motivation of occupying space would occasionally require expressing a temporary need for water in terms of the drive "thirst".
This drive in turn could generate one of several competing goals (choosing to which well to go).
After goal accomplishment, the need is satisfied and the drive is extinguished. 
However, the intrinsic motivation to keep occupying space would indefinitely persist, which will generate more needs, drives and goals downstream.

%(It feels weird, because if the objective never changes and never is fulfilled why then goals and drives are active "only to achieve the objective"? they will never succed on it), thus breaking the infinite regress. 

\subsection*{Theories of Intrinsic Motivation}

In standard theories of behavior, extrinsic reward maximization is the objective. Everything else, including intrinsic motivations, such as exploring or acting with curiosity, is a means to get even higher extrinsic reward in the future. 
In contrast, in theories of intrinsic motivation the reverse happens: movement, curiosity, and exploration -- ideas to be formalized next -- are the objective, while primary rewards are just a means to accomplish perpetual movement, broadly and abstractly understood.
 
Indeed, a common characteristic in intrinsic motivation theories is an irreducible desire to act and to move \cite{berridge2004motivation,di2017emerging,aubret2023information,rens2023evidence} -- either physically or in knowledge space. 
We take as an example the recently introduced maximum occupancy principle (MOP) \cite{moreno2023empowerment,ramirez2024complex} as a theory of intrinsic motivation, because other theories can be expressed as particular cases, or variations thereof. 
The main tenet of MOP is that agents should maximize future action-state path occupancy, which means that an agent should prefer performing actions leading to the visitation of the widest variety possible of future action and state sequences. 
These action-state paths can pertain to physical movement in the real world, but they can also correspond to neural activity paths that implement a diversity of cognitive processes, such as planning and thinking. 
Therefore, MOP proposes that maximizing the diversity and richness of physical and mental paths is the objective --the physical paths for conquering physical space, the mental ones for conquering knowledge space. 

\begin{figure}[htbp]
\centering
\includegraphics[width=1.0\textwidth]{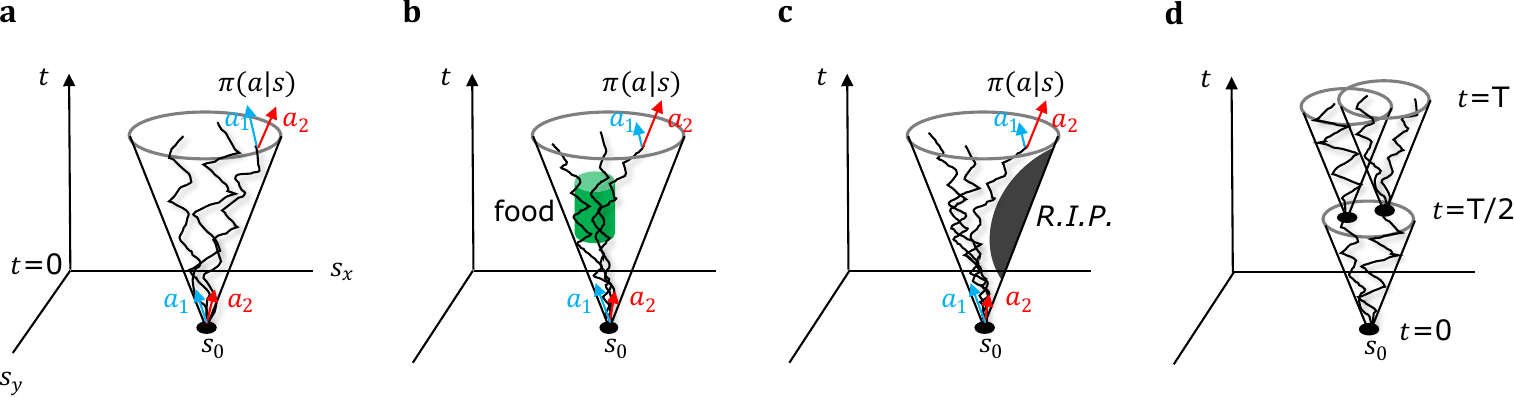}
\caption{
Intrinsic motivation in the form of maximizing action-state path entropy (a) generates diverse stochastic behavior, (b) shows goal-directed behavior when food sources (green region) are available, (c) makes the agent avoid terminal states (black region), and (d) provides a formalism that satisfies the additive property. 
Two initial and two final action pairs are indicated, sampled from policy $\pi(a|s)$, where $s=(s_x,s_y)$ is a two dimensional state variable in this particular case for illustration; actions are selected throughout the paths (not shown). Cones indicate the physically accessible region of action-state path space. 
In (d), path occupancy from time $t=0$ to time $t=T$ can be computed as the occupancy from time $t=0$ to $t=T/2$, plus the occupancy from $t=T/2$ to $t=T$ averaged over the intermediate action-states at time $t=T/2$, which fulfills the additive property of action-state path entropy, Eq. \ref{eq:entropy_additive}. This allows global optimization to be divided into subproblems.
Additivity also implies time-invariance (i.e., the future from $t=0$ looks the same as from $t=T/2$, everything else being the same), a fundamental law of nature. 
}
\label{fig:fig3}
\end{figure}

This idea of maximizing action-state path occupancy can be formalized as the objective of maximizing future action-state path entropy (ASPE); that is, maximizing the diversity of the paths that an agent can perform over action and state space  (Fig. \ref{fig:fig3}a; see Eq. \ref{eq:objective}). 
This amounts to roughly maximizing the logarithm of the number of paths over action-state space.
Some of the most powerful theories of intrinsic motivation, such as empowerment (MPOW, \cite{jung2011empowerment,brandle2023empowerment}), the free energy principle (FEP) \cite{friston2009reinforcement,dacosta2023reward,kiefer2025intrinsic}, information gain maximization (IGM, \cite{pathak2017curiosity,aubret2023information}) and the maximum occupancy principle (MOP, \cite{ramirez2024complex,moreno2023empowerment}) reduce to variations of this same idea -- MOP is directly related to the maximization action-state path entropy.
To make the idea of maximizing action-state path entropy more precise, we consider an MDP \((\mathcal{S}, \mathcal{A}, p, \gamma)\), where for simplicity \(\mathcal{S}\) is a discrete state space, and \(\mathcal{A}(s)\) is a state-dependent set of actions.

%Previous theories has contemplated entropy, KL and MI; the two terms can be combined, but MI cannot be combined because it is not additive; 

More formally, the agent's objective is to optimize the behavioral policy, denoted $\pi$, so as to maximize the ("motivation") value function
\begin{equation}
    V_{\pi}(s) = \sum_{t=0}^{\infty} \gamma^t  \mathbb{E}_{a_t \sim \pi, s_{t+1} \sim p} 
        \left[ \mathcal{H}(\pi(\cdot|s_t)) + \alpha \mathcal{H}(p(\cdot|s_t,a_t)) \bigg| s_0 = s \right]
        \ ,
        \label{eq:objective}
\end{equation}
where the expectation is over actions sampled from the policy and state transitions given initial condition $s$ at $t=0$, $\mathcal{H}(\pi(\cdot|s_t)) = - \sum_{a_t \in \mathcal{A}(s_t)} \pi(a_t|s_t) \log \pi(a_t|s_t)$ is the action entropy of a policy $\pi$ given state $s_t$, and $\mathcal{H}(p(\cdot|s_t,a_t))=- \sum_{s_{t+1}} p(s_{t+1}|s_t,a_t) \log p(s_{t+1}|s_t,a_t)$ is the state transition entropy conditioned on state $s_t$ and action performed $a_t$ at time $t$. 
The action entropy term encourages the agent to perform low-probability actions from any state rather than permanently relying on habitual behavior, thus favoring diversity of actions.
The state transition entropy pushes the agent towards more surprising state transitions.
Overall, the two terms promote action-state path diversity --as discussed below, the agent should also avoid the risks associated with this objective and sometimes use habitual or deterministic goal-directed behavior.
The discount factor $0 \leq \gamma<1$ measures how important future intrinsic motivation signals are compared to immediate ones. 
Because the value function considers future intrinsic motivation—not only immediate motivation—the problem becomes open-ended.
Also, the agent can trade off present for future action-state path entropy production, so that it can adopt deterministic behavior if doing so creates more future options.  
The parameter $\alpha$ weights the relative contribution of future state path entropy versus action path entropy, both of which contribute to the total value function $V(s)$.

The objective of maximizing action-state path entropy in Eq. \ref{eq:objective} generates diversity in action-state paths (Fig. \ref{fig:fig2}a-\ref{fig:fig3}). 
The outcome is an optimal policy that is stochastic, as opposed to being deterministic, effectively generating complexity in action-state paths.
Note that the signal $\mathcal{H}(\pi(\cdot|s_t)) + \alpha \mathcal{H}(p(\cdot|s_t,a_t))$ is a policy-dependent signal, which will be shaped by the evolution and learning of the policy itself; therefore this is an intrinsic reward (see Box 1).
The degree of produced diversity is also shaped by the state-dependence of the action set $\mathcal{A}(s)$. 
Indeed, most realistic environments have action bottlenecks (e.g., corridors) and, more importantly, terminal states (e.g., zero battery energy), where the size of available actions is reduced to just one action (e.g., staying still when dead or when having zero energy). 
State-dependent action sets strongly shape the resulting behavior, leading to maximum entropy policies that are non-uniform. In this case, the optimal policy promotes diverse action sequences while trying to avoid terminal states and bottlenecks \citep{ramirez2024complex}, unless bottlenecks lead to future large numbers of paths.

Action-state path entropy maximization produces behavior having the most basic features of naturalistic behavior (Fig. \ref{fig:fig3}a-d): (a) generation of complex, stochastic behavioral patterns; (b) goal-directed, pseudo-deterministic behaviors toward regions with food or energy sources, so that the agent can later diversify behavioral paths; (c) risk-avoidance and self-preserving behaviors when terminal states (dangerous, harmful or lethal conditions) are present or looming close; (d) rationality, as path occupancy can be computed additively and recursively only when using entropy measures \cite{ramirez2024complex,moreno2023empowerment}.
In particular, goal-directedness implies that the agent engages in a trade-off between immediate and future action-state path entropy, so that it can generate deterministic behavior in order to generate more stochasticity in the future.

Several theories of intrinsic motivation can readily be obtained from action-state path entropy maximization for particular choices of the parameters and interpretations of the state variables:

\begin{itemize}
\item Empowerment (MPOW) posits that agents should maximize the mutual information between future states and the action paths necessary to reach them \cite{jung2011empowerment,moreno2023empowerment}. While mutual information lacks the additive property \cite{ramirez2024complex} (Box 2), an approximation to mutual information maximization can be readily obtained by taking a negative contribution of the state transition entropy, $\alpha<0$, in Eq. \ref{eq:objective}. A negative parameter implies that predictable transitions have a higher value than unpredictable ones, and therefore actions leading to more deterministic ones will be preferred.
\item Information-gain maximization (IGM) proposes that agents should maximize the rate of mutual information increase between internal variables and the external world \cite{pathak2017curiosity,aubret2023information}. While this objective is again intractable due to lack of the additive property (Box 2), it can be approximated by Eq. \ref{eq:objective} when defining the state $s$ to represent knowledge state or belief, and choosing a large $\alpha \gg 1$.
This favors encountering surprising transitions from where the agent could learn; however, because of the presence of terminal states, only those surprising transitions that yield further surprising transitions would be favored. 
IGM is related to theories of novelty and surprise seeking \cite{becker2024representational}. 
\item The free energy principle (FEP) puts forward the idea that agents should minimize the surprise in their observations. Thus, agents should counteract perturbations pushing them away from desired homeostatic states (e.g., appropriate body temperature range). 
Formally, FEP minimizes the Kullback-Leibler (KL) divergence between the actual distribution $p(s)$ over states and the desired/target distribution $q(s)$ \cite{friston2009reinforcement,dacosta2023reward}. FEP is a variation of Eq.  \ref{eq:objective} where $\alpha \gg 1$ with an added extrinsic reward function term $ \alpha \log q(s_{t+1}) $ (see Box 1). Because of this construction, FEP does not contain an action entropy term; instead, there is an added term that strongly penalizes (negative extrinsic reward) for not being close to the target distribution.
%production and there is a big penalty (negative reward) for not being close to the target distribution. 
\item The maximum occupancy principle (MOP) is identical to action-state path entropy maximization in Eq. \ref{eq:objective} \cite{ramirez2024complex,moreno2023empowerment}. While a natural choice is $\alpha=1$, several variations of MOP have already been studied where $\alpha$ is taken to be zero, so that only action path entropy generation is maximized \cite{mastrogiuseppe2024controlled,mastrogiuseppeplayful,habibunsupervised}. 
\end{itemize}

Additional intrinsic motivation objectives can be obtained under further limits and approximations. For instance, causal entropic forces \cite{wissner2013causal} can be obtained from Eq. \ref{eq:objective} by setting a finite, fixed horizon over which to compute the value function. Under such a finite-horizon setting, this intrinsic motivation reduces to maximizing the entropy of the final state, rather than the entropy over entire state-action paths.
This amounts to taking a single term of the sum in Eq. \ref{eq:objective} and considering large $\alpha \gg 1$. 
%The Free Energy approach \cite{friston2009reinforcement,dacosta2023reward} can be obtained from Eq. \ref{eq:objective} with a large $\alpha \gg 0$, so that only state transitions are considered, and adding a set of desired, or target, states in the form of extrinsic reward $\log q(s_{t+1}|s_t,a_t)$, so that agents create a tendency to move towards those target states -- this leads to a Kullback-Leibrer (KL) objective \cite{dacosta2023reward} shown to by minimized by a deterministic policy \cite{moreno2023empowerment}.
An approximation of action-state path entropy maximization can be used for the generation of self-organized collective motion \cite{devereux2023environmental} by taking the limit $\gamma \rightarrow 1$ and $\alpha \gg 1$ in Eq. \ref{eq:objective}, so that only the entropy of the long-run state visitation distribution is considered.

While many intrinsic motivation approaches are based on mutual information \cite{jung2011empowerment,mohamed2015variational,sharma2019dynamics}, the use of entropy instead offers the advantage that the optimization problem admits a recursive Bellman formulation because entropy accumulates additively in MDPs due to conditional independence of policies and state transitions (Fig. \ref{fig:fig3}d). 
Beyond the simplicity gained by having a recursion, additivity is important because it enables time-invariant decision policies, which align with principles of stationarity found in physical laws.
%additivity is important because it is equivalent to assuming time-invariance of the decision policies, a cornerstone in all laws of physics. 
We leave a deeper discussion of this for Box 2. 

%In reinforcement learning theory and applications, intrinsic motivation of the above type are used in \cite{mohamed2015variational,grytskyy2023general,pathak2017curiosity}. Many approaches originally use maximization of mutual information \cite{jung2011empowerment,mohamed2015variational,sharma2019dynamics}, but this requires variational approximations, and the objective does not satisfy the additive property that is required to enforce time homogeneity as a principle of rational behavior (see Box 2).  

\subsection*{Intrinsic Motivation within the Confines of Embodiment and Bounded Cognition}

Theories of extrinsic reward maximization predict that behavior collapses to a deterministic behavioral policy after enough learning \cite{sutton1998reinforcement} (see Fig. \ref{fig:fig2}b).
In contrast, natural behavior is characterized as being wide and complex, showing both large variability \cite{renart2014variability,urai2025structure} and diverse patterns.
Indeed, behavioral collapse to repetitive patterns is typically considered signs of mental disorders, such rumination and negative thinking in depression and eating disorders \cite{palmieri2021repetitive,rickerby2024rumination}.
In the previous section we have introduced intrinsic motivation as the main force that expands behavior to generate complex patterns.
Intrinsic motivation generates complex behaviors--ideally, the full range of behaviors that the agent is capable of. 
Yet there should be constraints that ultimately shape behavior. 
%Now we describe the forces that limit behavior. These are better understood as bounds/boundaries, rather than tendencies or forces.
We describe these not as driving forces or tendencies, but as bounds or boundary conditions that constrain behavior.
These bounds take the form of resource-limited cognition and embodiment. 

Resource-limited, or bounded, cognition \cite{russell1991principles,ortega2015information,moreno2020heuristics} is a theory proposing that agents strive to maximize reward under limited resources, including limited memory, time or energy.
Even though behavior could be in principle more complex, the limitation, for instance, to remember few items in working memory, shapes task solving, a constraint that can be mitigated through extended cognition \cite{sutton2010psychology}.
Similarly, the impossibility of contemplating all potential consequences limits our planning capabilities \cite{simon1972theories,mastrogiuseppe2022deep}, again shaping the complexity of behaviors that can be generated. 

Perhaps, however, the major limitation of behavior comes from the agent's physical instantiation -- a concept known as embodiment. For instance, thumb-to-fingers opposition in both rodents and primates allows us to easily grasp and manipulate small objects, while birds are less dexterous because they lack this opposition \cite{sugasawa2021object}.
Systematic biases in decision making can be readily interpretable as motor constraints that need to be considered during planning \cite{cos2011influence,fievez2024task}.
Recently, the dynamics of the body have been integrated with increasing degree of realism into brain models to better predict behavior in the fly \cite{vaxenburg2024whole} and in rodents \cite{aldarondo2024virtual}. 
Realism in body dynamics and physical environment allows studying embodied cognition. 
For instance, the body limitations of the 4-leg Ant (an artificial agent in the MuJoCo physics engine to model realistic locomotion and surface contact physics in a quadruped \cite{todorov2012mujoco}; Fig. \ref{fig:fig2}a) do not allow it to fly and forbids many other movements. Nevertheless, its body and contact physics does not preclude it spinning and jumping by flexing and then rapidly extending several legs to propel itself.
%by propelling itself after flexing and rapidly extending several legs
These behaviors naturally arise in combination with, for instance, the MOP intrinsic motivation \cite{ramirez2024complex} (\href{https://drive.google.com/file/d/1hkXX1mfZOlGB_EnYgthkfD54WdyeVjQW/view?usp=drive_link}{Video 1}).
The integration of intrinsic motivation within embodied agents promises to yield a more complete understanding and better tools for the modeling of natural behavior.  

Among the most important constraints imposed by the body is the presence of {\em terminal} states--that is, states that lead to life-threatening damage or disintegration of the agent. 
These include dehydration and hypoglycemia, involving severe homeostatic imbalances \cite{vergara2019energy}, but also purely bodily damage, such as devastating broken bones or concussions.
As reaching a terminal state implies being able to produce little or no action-state entropy thereafter, natural agents must develop behavioral strategies to avoid them.
For instance, to avoid dehydration, a thirst sensation and an automatic drive to drink constitute a warning that signals the proximity to a terminal state. 
Similarly, while action-state entropy would promote the generation of jerky and highly random limb movements, high frequency movements risks causing the deterioration and damage of bone joints and ligaments.
Therefore, warning signals like muscle pain and fatigue might have evolved as mechanisms to avoid reaching too close to those terminal states, setting strong constraints on the limits that intrinsic motivation can reach.   

One might wonder whether it is possible to map terminal states into an extrinsic reward function. For instance, one could associate an infinite negative extrinsic reward to terminal states. However, the effect on behavior would not be the same: the result of having states with infinite negative reward would be to produce a deterministic policy, one that pushes the agent very far from those terminal states. In contrast, an intrinsic motivation like MOP would make agents occupy path space, while explicitly excluding transitions to terminal states. The result is thus a stochastic policy even in deterministic environments. In summary, an extrinsic reward leads to maximal avoidance, while intrinsic motivation leads to maximal safe path diversity.
%Terminal states are not the same as having infinite reward over those states, a possible way to try to map intrinsic reward function into extrinsic ones. This is because setting negative extrinsic reward to the terminal states makes the optimal policy to be deterministic, while the optimal policy under intrinsic motivation is typically stochastic, pursuing variability in itself. 

\subsection*{Intrinsic Motivation Signals in the Brain} 

Neuronal activity across multiple brain regions is linked to the acquisition and consumption of primary rewards, such as food and water, as well as to the processing of secondary rewards—cues that predict those primary outcomes \cite{schultz2000multiple}. These neural signals can serve as reinforcers that shape goal-directed behavior. Such behavior, however, depends on integrating external rewards with internal physiological states, since the value of a reward like food varies with factors such as satiation \cite{araujo2020rethinking}.
%Neuronal signals from several areas across the brain correlate with the acquisition and consumption of primary rewards, such as water and food, as well as the perception of secondary rewards (i.e., cues predicting primary rewards) \cite{schultz2000multiple}. These signals can act as reinforcers, driving goal-directed behavior. However, goal-directed behavior requires the integration of external rewards with internal states, as the value of food, for instance, depends on satiation \cite{araujo2020rethinking}. 

A better understanding of how extrinsic and intrinsic rewards are integrated to guide behavior has emerged, yet the underlying mechanisms appear intricate. For instance, dopaminergic neurons in the substantia nigra and ventral tegmental area encode not only reward prediction errors \cite{schultz2000multiple} but also reward uncertainty \cite{fiorillo2003discrete}. This blending of prediction-error and uncertainty signals aligns with intrinsic motivation theories, which incorporate uncertainty and exploration terms such as $-\log p(s_{t+1}\mid s_t, a_t)$ in Eq. \ref{eq:objective}.
%How extrinsic and intrinsic rewards are integrated to orchestrate behavior has started to be elucidated, but seems intricate. For example, dopaminergic neuronal activity in substantia nigra and ventral tegmental area \cite{schultz2000multiple} represent reward prediction errors, but also reward uncertainty \cite{fiorillo2003discrete}. Mixing signals about prediction errors and uncertainty is a feature predicted by intrinsic motivation theories (see terms $-\log p(s_{t+1}|s_t,a_t)$ in Eq. \ref{eq:objective}).

Studies of mice exploring open arenas without extrinsic rewards \cite{markowitz2023spontaneous} have shown that dopamine is spontaneously released in the dorsolateral striatum, randomly coinciding with behavioral syllables (e.g., rearing or running). When a dopamine transient aligns with a particular syllable, the probability of that behavior being selected again increases, and the variability (entropy) of subsequent syllable sequences also rises. These findings suggest that spontaneous dopamine release may encode the “surprise" of forthcoming actions or syllables, which can be modeled as an intrinsic reward $r_t = -\log \pi(a_t \mid s_t)$ (see Eq. \ref{eq:objective}),  where an action is more rewarded if it is unlikely under the current behavioral policy.
%Work in mice moving in open arenas with no extrinsic reward \cite{markowitz2023spontaneous} has found that dopamine is spontaneously released in the animal's dorsolateral striatum, randomly coinciding with behavioral syllables (such as rearing or running). It is observed that the temporal association between spontaneous dopamine release and a coinciding behavioral syllable increases the likelihood of that behavior being chosen in the future and enhances the variability (entropy) of future syllable sequences. Therefore, spontaneous dopamine release may signal the prediction of future action or syllable "surprise" $r_t=- \log \pi(a_t|s_t)$ (see Eq. \ref{eq:objective}), which serves as a central signal in some intrinsic motivation frameworks. 

State-transition surprise signals of the form $r_t = -\log p(s_{t+1}\mid s_t, a_t)$ may also be reflected in dopamine release. Although dopamine ramps are typically interpreted as deviations of reward-prediction–error signals \cite{kim2020unified}, it remains possible that surprising state transitions modulate dopamine levels—either increasing or decreasing release—depending on the animal’s proximity to reward. The influence of state-transition surprise is likely nuanced: large or abrupt transitions may be interpreted by rodents as trial abortion or even as a potential threat, such as a sign of predators. In line with this interpretation, dopamine release has been observed to rise when rodents retreat from a novel object that is likely perceived as threatening \cite{akiti2022striatal}.
%State transition surprise signals of the form $r_t -\log p(s_{t+1}|s_t,a_t)$ might also correlate with dopamine release. Although dopamine ramps are currently explained as extensions of reward prediction errors due to perturbations \cite{kim2020unified}, there is the possibility that surprising state transitions can either increase or reduce dopamine release, depending on the distance to reward. Indeed, the overall effect of surprise in state transitions can be complex, as large state transitions might be perceived by rodents simply as trial abortion or as threats, such as proximity to predators. Consistent with this, dopamine release increases when rodents retreat from a novel object probably considered as threatening \cite{akiti2022striatal}.

Studies in non-human primates have shown that they actively work to obtain information, even when that information cannot change a future outcome. Their brains treat this information as an intrinsic reward, with midbrain dopamine neurons firing in response to informative cues just as they would for a drop of juice \cite{bromberg2009midbrain}. Moreover, neurons in the prefrontal cortex, orbitofrontal cortex in particular, calculate the specific value of this information, allowing the monkeys to make sophisticated economic decisions about whether to "pay" to resolve their uncertainty \cite{blanchard2015monkeys}. Finally, novelty can provide an intrinsic bonus and enhance salience signals in the posterior parietal cortex independently of reward \cite{foley2014novelty}. 

Notably, the zona incerta (ZI), a small structure inside the subthalamus, has been demonstrated to be critically involved in novelty seeking and exploration behavior in both mice and monkeys \cite{ahmadlou2021cell,ogasawara2021neuronal}. The ZI's extensive connections with motor centers in the basal ganglia and the brainstem, alongside its connections to the thalamus, a key region for attention and sensory gating, form a potential circuitry for driving these behaviors.

\subsection*{Concluding Remarks and Future Perspectives}

Theories of intrinsic motivation posit that intrinsic reward signals are more fundamental than extrinsic reward signals to orchestrate behavior. 
These theories, in one way or another, propose that action-state path entropy production is the overarching objective under which all behavior can be understood.
Drives, goals and extrinsic reward seeking are a means to accomplish the objective, not the objective by themselves; they are temporary and transient, while the objective of action-state path maximization is always active. 
Intrinsic motivation ensures a diversity of behaviors while accounting for extrinsic rewards, embodiment, bounded cognition and terminal states.
%The result is open-ended embodied behavior 
An appealing aspect of intrinsic motivation is that, in principle, it can generate all behaviors the agent is capable of while considering its limited cognitive resources and embodied situatedness.  
Generative models are popular in text \citep{brown2020language}, image \citep{goodfellow2014generative}, and video generation \citep{vondrick2016generating}, but they are largely lacking in behavioral sciences.
Therefore, intrinsic motivation theories provide a unique opportunity to build generative models of open-ended, embodied behavior, by going beyond extrinsic reward maximization and task-dependent optimization.

\subsection*{Outstanding Questions}

\begin{itemize}
    \item How to distinguish between extrinsic and intrinsic reward signals across different brain areas? 

    \item How to integrate brain and body models? How to scale them up? 

    \item How to arbitrate between different theories of intrinsic motivation? Qualitative differences should be extracted from each theory, though environmental noise could obscure distinguishability. Paradoxical limiting cases could guide experimental design and model comparison (Box 3).

    \item How to build large-scale models of behavior with intrinsically motivated embodied multi-agents? 
\end{itemize}

\newpage
% FIGURES

\newpage

\begin{theorybox}[Extrinsic and Intrinsic Rewards]

\vspace{0.2cm}

In reinforcement learning \cite{sutton1998reinforcement}, the objective is to optimize the policy $\pi(a|s)$ so as to maximize the value function
\begin{equation}
    V_{\pi}(s) = \sum_{t=0}^{\infty} \gamma^t  \mathbb{E}_{a_t \sim \pi, s_{t+1} \sim p} \; r(s_t,a_t,\pi(a_t|s_t))  \ ,
        \label{eq:objective_reward}
\end{equation}
where $r(s,a,\pi(a|s))$ is a reward function depending on the current state and action performed by the agent, and as well as of the policy.
This optimization amounts to finding the policy $\pi$ that maximizes the cumulative sum of expected future rewards starting from state $s$.
Depending on whether the reward function includes or not the policy, we define:

\begin{itemize}
   
\item \textbf{Extrinsic Reward}: A reward is extrinsic if it is a policy-independent, state-action $(s,a)$ signal sought by an agent, that is, a function $r(s,a)$ only, independent of $\pi$. 
An extrinsic reward can be an external or internal signal sought by an agent.
Examples are sugar in the mouth and electrical stimulation in nucleus accumbens. 
The nature of extrinsic reward cannot be changed by the agent, as it is fixed by design or nature. The extrinsic reward function $r(s,a)$ is sometimes called utility, and very often it is simply referred as reward.

\vspace{0.2cm}

\item \textbf{Intrinsic Reward}: Signal sought by an agent that is policy-dependent, that, is a function $r(s,a,\pi)$ that explicitly depends on the policy. The nature of an intrinsic reward can be changed by the agent, as it depends on its own behavioral policy. One example is empowerment or the maximum occupancy principle, both of which use the entropy of sequences of actions as an integral component. One specific example of intrinsic reward function is $r(s,a,\pi) = -\log \pi(a|s)$ for $\alpha = 0$ in Eq. \ref{eq:objective}. In this review we have equated intrinsic reward with intrinsic motivation. The central assumption is that an extrinsic reward function, which is fixed by construction, cannot model an intrinsic motivation. 

\end{itemize}

Extrinsic reward signals are difficult to design in robotic applications and are typically sparse, so that learning based on them can be slow. In contrast, intrinsic rewards generate signals useful for many tasks, such as movement and exploration generation; further, they provide denser reward signals, which can help mitigate sparse reward problems and accelerate learning.
%they are dense, and therefore useful for accelerating learning, as gradients are almost never vanishing. 
For instance, the intrinsic reward signal $r(s,a,\pi) = -\log \pi(a|s)$ fosters exploration and generation of complex behavioral patterns. 
%Additional teaching signals arise from terminal states and warning states surrounding them; reaching a warning state (such as falling) provide relatively frequent feedback signals in natural conditions during learning. 

\end{theorybox}

\newpage

\begin{theorybox}[Entropy or Mutual Information, What Is More Fundamental?]

\vspace{0.2cm}

What is more fundamental: the notion of entropy or the notion of information?
The original formulation of MPOW and IMG formulate their objectives using some measure of mutual information, while MOP and FEP focus on entropy and KL, respectively. Is one measure of intrinsic reward more fundamental than the other? And what should this choice be based on? 
One possibility is to choose the measure that is more tractable from the computational standpoint, and/or more rational. For instance, if a measure can be written as the sum of per-step contributions, then any problem could be easily broken into smaller parts that could be solved greedily and individually. However, if a measure does not lend itself to this division, then it might be harder to optimize and generalize from it. Also, the former property, called additivity (see Fig. \ref{fig:fig3}d), would allow the problem to be naturally expressed in the form of a Bellman equation, a recursion that connects values of current states with future state values. Furthermore, additivity is equivalent to assuming that time is homogeneous, that is, no time is different to any other time; building an objective that is additive therefore also reflects a form of rationality. 

\vspace{0.2cm}

It can be shown that whereas action-state path entropy is additive in MDPs, mutual information (MI) between action and state paths is not. 
This means that MI is not additive over paths: the information between a state path $s=(s_1,...,s_t,...,s_T)$ and the action path $a=(a_0,...,a_{t-1},...,a_{T-1})$ that generated the former is not the same as the sum of the information of between early subpaths $s_{A}=(s_1,...,s_t)$ and $a_{A}=(a_0,...,a_{t-1})$ and later subpaths $s_{B}=(s_{t+1},...,s_T)$ and $a_{B}=(a_{t},...,a_{T-1})$.
More formally, the total mutual information is not the sum of its parts \cite{ramirez2024complex,moreno2023empowerment}, that is,
\begin{equation}
     \text{MI}(s,a) \neq \text{MI}(s_A,a_A) + \text{MI}(s_B,a_B|s_t,a_{t-1}) \ 
        \label{eq:MI_non_additive}
\end{equation}
 -- an exception occurs when the action and state paths are independent, leading to zero MI. 
Intuitively, and in general, there is additional information between the early and late paths that is not captured only by the conditioning on the intermediate point $(s_t,a_{t-1})$ in the second term of the equation.
In contrast, the entropy of the action-state paths is the sum of the entropies of any subpaths, averaged over intermediate points (see Fig. \ref{fig:fig3}d), that is \cite{ramirez2024complex,moreno2023empowerment}, 
\begin{equation}
    \mathcal{H}(s,a) = \mathcal{H}(s_A,a_A) + \mathcal{H}(s_B,a_B|s_t,a_{t-1}) \ .
        \label{eq:entropy_additive}
\end{equation}

\vspace{0.2cm}

The non-additivity of MI precludes it from being written in the form of a Bellman equation, while action-state path entropy (and thus MOP, FEP and KL) is amenable to Bellman recursions. The Bellman recursion ensures that entropy is well-suited for open-ended behavior, as there is no need to define episodes with arbitrary horizons.
%as it is mandatory in MI approaches due to its non-additivity. 

\end{theorybox}

\newpage

\begin{theorybox}[Paradoxes of Intrinsic Motivation]

\vspace{0.5cm}

{\rowcolors{3}{gray!10!}{gray!30}
\begin{tabular}{ |p{7cm}|p{5cm}|  }
\hline
\multicolumn{2}{|c|}{ {\em Paradoxes of Intrinsic Motivation} } \\
\hline
\rowcolor{gray}
\textcolor{white}{Theory} & \textcolor{white}{Paradox} \\
\hline
Extrinsic Reward Maximization &	Buridan’s Donkey \\
FEP	& Dark Room \\
MOP	& Grand Roulette \\
IGM	& Noisy TV \\
\hline
\end{tabular}}

\vspace{0.5cm}

Theories of behavior are sometimes poisoned with paradoxes. Paradoxes may not be realizable in practice, but they can inform about the behavior expected in limit conditions. Therefore, they are not only of theoretical interest but also of practical significance.
%they do not only have theoretical interest, but they also hold practical significance. 
Among the most relevant paradoxes we find:

\begin{itemize}

\item Extrinsic reward maximization theories face the {\em Buridan’s donkey} paradox. In this imaginary experiment, an extremely and equally hungry and thirsty donkey is placed exactly between a hay and water bales. As the donkey remains undecided, it starves to death. 

\vspace{0.2cm}

\item FEP encounters the {\em dark room} \cite{friston2009reinforcement}. A person whose objective is to minimize the divergence between actual and homeostatic states will be willing to confine themselves to a dark room containing all needed to maintain homeostasis (water, food and shelter). The minimal distance between actual and desired states will be obtained, precisely, inside the dark room, so that they will have no incentive to leave it.  

\vspace{0.2cm}

\item MOP faces the {\em grand roulette}. A gambler in a casino, encountering a grand roulette, with a myriad of possible bets, will be trapped to play to it ad infinitum. This will generate numerous action state paths, which fulfills their desire to maximize action-state path entropy.  

\vspace{0.2cm}

\item IGM will cause a person to be trapped by a {\em noisy TV} \cite{burda2018large}. If the person does not know that the TV is actually noisy and that there is nothing to learn, not even hidden messages, they will be willing to watch it to try to maximize information gain forever. 

\end{itemize}

Although these paradoxes caricature the intrinsic motivation approaches, they also highlight crucial differences between them: while some promote controllability and homeostasis, others foster behavioral diversity and curiosity.

\end{theorybox}

\newpage

\subsection*{Acknowledgments}
This project was supported by grants funded by the Spanish Ministry of Science, Innovation and Universities (MICIU/AEI/10.13039/501100011033) and by “FEDER A way of making Europe” (ref: PID2023-146524NB), and by ICREA ACADÈMIA (2022) funded by the Catalan Institution for Research and Advanced Studies to R.M.B.

%%%  Use this section to acknowledge contributions 
%%%  from non-authors and list funding sources, 
%%%  including grant numbers.

\newpage

\bibliography{references}

\end{document}